\documentclass[sigconf]{acmart}

\usepackage{mathtools}
\usepackage{graphicx}
\usepackage{courier}
\usepackage{epstopdf}
\usepackage{color}
\usepackage{csquotes}

\usepackage{amsmath}
\usepackage{amsfonts}
\usepackage{amssymb}
\usepackage[subnum]{cases}
\usepackage{booktabs} 
\usepackage{pifont}
\usepackage{graphics}

\usepackage{amssymb}
\usepackage{pifont}

\usepackage {tikz}
\usetikzlibrary {positioning}
\definecolor {processblue}{cmyk}{0.96,0,0,0}
\usepackage{lipsum,adjustbox}

\usepackage[subnum]{cases}
\usepackage{color,soul}

\usepackage{algorithm,algorithmicx,algpseudocode}

\usepackage{subcaption}

\usepackage{relsize}

\newtheorem{lemma}{Lemma}

\usepackage{footnote}
\makesavenoteenv{tabular}

\usepackage{multirow}

\usepackage{amsmath, amssymb, bbm, xspace}

\def\spose#1{\hbox to 0pt{#1\hss}}

\def\text #1{\hbox{\quad#1\quad}}

\def\nthinsp{\mskip -2   mu}

\def\superstar{^{\raise 0.5pt\hbox{$\nthinsp *$}}}
\def\SUPERSTAR{^{\raise 0.5pt\hbox{$*$}}}

\def\lamstarT {\lambda^{\raise 0.5pt\hbox{$\nthinsp *$}T}}

\def\hbar{\skew{4.2}\bar h}

		\def\bk1{{\rm 1\kern-.17em l}}
		\def\bkD{{\rm I\kern-.17em D}}
		\def\bkR{{\rm I\kern-.17em R}}
		\def\bkP{{\rm I\kern-.17em P}}
		\def\bkY{{\bf \kern-.17em Y}}
		\def\bkZ{{\bf \kern-.17em Z}}

		\def\beq{\begin{eqnarray}}
		\def\bc{\begin{center}}
		\def\be{\begin{enumerate}}
		\def\bi{\begin{itemize}}
		\def\bs{\begin{small}}
		\def\bS{\begin{slide}}
		\def\ec{\end{center}}
		\def\ee{\end{enumerate}}
		\def\ei{\end{itemize}}
		\def\es{\end{small}}
		\def\eS{\end{slide}}
		\def\eeq{\end{eqnarray}}
		\def\qed{\quad \vrule height7.5pt width4.17pt depth0pt}

	\def\cp2problem#1#2#3#4{\fbox
		 {\begin{tabular*}{0.9\textwidth}
			{@{}l@{\extracolsep{\fill}}l@{\extracolsep{6pt}}l@{\extracolsep{\fill}}c@{}}
				#1 & & $#4 $
			\end{tabular*}}}

		\renewcommand{\emph}[1]{\textbf{#1}}

		\def\bk1{{\rm 1\kern-.17em l}}
		\def\bkD{{\rm I\kern-.17em D}}
		\def\bkR{{\rm I\kern-.17em R}}
		\def\bkP{{\rm I\kern-.17em P}}
		
		\def\bkZ{{\bf{Z}}}

\newcommand {\beeq}[1]{\begin{equation}\label{#1}}
\newcommand {\eeeq}{\end{equation}}
\newcommand {\bea}{\begin{eqnarray}}
\newcommand {\eea}{\end{eqnarray}}

\def\texitem#1{\par\smallskip\noindent\hangindent 25pt
               \hbox to 25pt {\hss #1 ~}\ignorespaces}

\usepackage{hyperref}
\usepackage{cleveref}[2012/02/15]

\crefformat{footnote}{#2\footnotemark[#1]#3}

\setcopyright{rightsretained}

\acmDOI{10.475/123_4}

\acmISBN{123-4567-24-567/08/06}

\acmConference[Conf'18]{}{XXXX}{XXXX} 
\acmYear{1997}
\copyrightyear{2016}

\acmArticle{4}
\acmPrice{15.00}

\begin{document}
\title[SaC2Vec: Network Embedding]{SaC2Vec: Information Network Representation with Structure and Content}

\author{Sambaran Bandyopadhyay}
\affiliation{\institution{IBM Research}}
\email{sambband@in.ibm.com}

\author{Harsh Kara}
\affiliation{\institution{Indian Institute of Science, Bangalore}}
\email{harshk@iisc.ac.in}

\author{Anirban Biswas}
\affiliation{\institution{Indian Institute of Science, Bangalore}}
\email{anirbanb@iisc.ac.in}

\author{M N Murty}
\affiliation{\institution{Indian Institute of Science, Bangalore}}
\email{mnm@iisc.ac.in}




\begin{abstract}
Network representation learning (also known as information network embedding) has been the central piece of research in social and information network analysis for the last couple of years. An information network can be viewed as a linked structure of a set of entities. A set of linked web pages and documents, a set of users in a social network are common examples of information network. Network embedding learns low dimensional representations of the nodes, which can further be used for downstream network mining applications such as community detection or node clustering. Information network representation techniques traditionally use only the link structure of the network. But in real world networks, nodes come with additional content such as textual descriptions or associated images. This content is semantically correlated with the network structure and hence using the content along with the topological structure of the network can facilitate the overall network representation.

In this paper, we propose \textit{Sac2Vec}, a network representation technique that exploits both the structure and content. We convert the network into a multi-layered graph and use random walk and language modeling technique to generate the embedding of the nodes. Our approach is simple and computationally fast, yet able to use the content as a complement to structure and vice-versa. We also generalize the approach for networks having multiple types of content in each node. Experimental evaluations on four real world publicly available datasets show the merit of our approach compared to state-of-the-art algorithms in the domain.
\end{abstract}

\settopmatter{printacmref=false} 

\maketitle

\section{Introduction}\label{sec:intro}

Mining information networks is important for different research and business purposes. Community detection for product advertisement in commercial network, paper classification in a citation network are some of the important applications. But traditional machine learning and data mining algorithms suffer in these applications because of the very high dimension (millions of nodes can be there) and extreme sparsity (a node is directly connected only to a very small subset of nodes) in the large network. So the success of the mining tasks depend on the efficient feature selection and design of the network. Typically efficient feature engineering needs a lot of domain knowledge, experiments and human efforts. 
Compared to that, in a network embedding framework \cite{perozzi2014deepwalk}, a function to represent each node in the form of a compact and dense vector is learnt mostly in a task independent way. As shown in the literature \cite{grover2016node2vec,tang2015line}, machine learning algorithms perform better on these embeddings for different network mining tasks such as node classification, link prediction, etc.

There has been significant improvement in the network embedding literature in the last few years. But popular information network representation techniques, such as DeepWalk \cite{perozzi2014deepwalk}, node2vec \cite{grover2016node2vec}, struc2vec \cite{ribeiro2017struc2vec} use only the link structure of the network to find the node embedding. But most of the real world networks come also with rich content associated with each node. They can complement the structural information, specially when the structure is noisy. For example, there are millions of active and connected users in Twitter and each of them can post thousands on tweets \cite{huang2017accelerated}. Content of these tweets can include text, images or videos. It is natural that connection between the users in such social networks depends on the content that they post. So to understand the underlying semantics of a network, it is important to consider this rich content information along with structure to model the overall network. Sociological theories such as homophily \cite{mcpherson2001birds} also suggest a strong correlation between the structure and the content of a network. Besides, content has been shown to be useful for some mining tasks such as tackling filter bubble problem in networks \cite{DBLP:conf/wsdm/LahotiGG18}, evolving social action prediction \cite{tan2010social}, etc. Hence combining content with structure for information network representation can help the node embeddings to be topologically and semantically coherent.

Integrating content of each node into the state-of-the-art network embedding techniques is challenging. However there are few matrix factorization based approaches \cite{yang2015network,huang2017accelerated} present in the literature where content or the node attributes have been used with topological structures to generate the node embeddings. Deep learning based and semi-supervised approaches have also been proposed recently to model structure with other attributes \cite{hamilton2017inductive,hamilton2017inductive,huang2017label} for network embeddings. These approaches have some limitation in the sense that they often need heavy computing power and memory or good amount of supervision to produce the desired results. Hence in this paper, we motivate the use of simple, fast and efficient methods to combine content along with structure for network representation learning. We first give two intuitive approaches for this task, and finally propose a novel technique \textit{SaC2Vec} (structure and content to vector) which uses both structure and content in an intelligent way to find the embedding for each node in the network. It creates a multi-layer network and employs random walk along with language modeling techniques to generate the lower dimensional representation of the nodes.\\

\textbf{Contributions}: Following are our contributions in this paper:
\begin{itemize}
\item We motivate the use of content along with the structure for information network embedding. We propose some simple and intuitive approaches first, and also demonstrate their usefulness through experiments.
\item We propose a novel unsupervised algorithm SaC2Vec, which creates a multi-layer graph and employs random walk along with language modeling techniques to generate the lower dimensional representation of the nodes. To the best of our knowledge, this is the first network embedding technique which employs a multi-layered graph where one layer corresponds to structure and each of the other layers deals with different types of content.
\item We evaluate the performance of all the proposed algorithms on medium to large size datasets for different types of mining tasks such as node classification, node clustering and network visualization. The results are compared with different state-of-the-art embedding algorithms to show the merit of our approach.
\end{itemize}


\section{Related Work}\label{sec:related}
We summarize the existing literature in this section, and find the potential research gap to be addressed in this work. Detailed survey on Network Representation Learning (NRL) can be found in \cite{hamilton2017representation} or \cite{zhang2017network}. 
There has been a lot of work on feature engineering for networks. Most of the earlier work in this area tried to design hand-crafted features based on domain expertise in networks \cite{gallagher2010leveraging}. Different efficient dimensionality reduction techniques were used to generate lower dimensional network representations. Various linear and non-linear unsupervised approaches such as PCA \cite{wold1987principal} and ISOMAP \cite{bengio2004out} have also been used to map the network to a lower dimensional vector space. 

In recent times, representation learning in the domain of natural languages exhibited promising improvement. This has motivated the researchers to use those ideas and adopt them to be used for information network embedding. The idea of representing words \cite{mikolov2013efficient} in a document in the form of an embedding vector was used in DeepWalk \cite{perozzi2014deepwalk} to represent nodes in networks. One can find a one-to-one correspondence between a word in a corpus and a node in a network. DeepWalk employs a uniform random walk from a node until the maximum length of the random walk is reached. 
Line \cite{tang2015line} uses two different optimization formulations to explicitly capture the first order and second order proximities in node embeddings. It uses edge sampling strategy to efficiently optimize the objectives.
In node2vec \cite{grover2016node2vec}, a biased random walk is proposed to balance between the breadth first search and depth first search while generating the corpus via random walk. Subsequently it also uses language models to find the node embeddings in a network.
Struc2vec \cite{ribeiro2017struc2vec} is another random walk based node embedding strategy which finds similar embeddings for the nodes which are structurally similar.
Social rank has been considered with the higher order proximities to generate node embeddings in \cite{gu2018rare}.
These methods are fast and efficient but they consider only the link structure of a network.

Nonnegative matrix factorization based network embedding techniques have been popular for network embedding. A network representation approach based on factorizing higher orders of adjacency matrix has been proposed in \cite{cao2015grarep}. In \cite{tu2016max}, authors have shown the equivalence of DeepWalk embedding technique to that of a matrix factorization objective, and further combined that objective to a max-margin classifier to propose a semi-supervised network embedding technique. Modularity maximization based community detection method has been integrated with the objective of nonnegative matrix factorization in \cite{wang2017community} to represent an information network. In \cite{yang2017fast}, authors have captured different node proximities in a network by respective powers of the adjacency matrix and proposed an algorithm to approximate the higher order proximities. 

Deep learning based network embedding techniques are also present in the literature. In \cite{wang2016structural}, authors propose a structural deep network embedding method. They first propose a semi-supervised deep model with multiple layers of non-linear functions. Then second-order proximity is used by the unsupervised component to capture the global network structure. The idea of using convolutional neural networks for graph embedding has been proposed in \cite{niepert2016learning}, and further developed in \cite{hamilton2017inductive} where authors propose GraphSAGE which learns the network embedding with node attributes in an inductive setting. Heterogeneous network embedding using a deep learning network is proposed in \cite{chang2015heterogeneous}.
CANE \cite{tu2017cane} learns multiple embeddings for a vertex according to its different contexts.

A major limitation in all of the above works is that they use only the network structure for embedding. But for most of the real-world networks, rich content information such as textual description is associated with the nodes. Integrating such content is not straightforward in any of the above approaches. There is some amount of work present in the literature to combine structure with content for network representation. In \cite{yang2015network}, authors have presented a matrix factorization based approach (TADW) for fusing content and structure. 
An attributed network embedding technique AANE is proposed in \cite{huang2017accelerated}. The authors again uses matrix factorization to get low dimensional representation from the attribute similarity matrix, and use link structure to maintain the network proximity in the embedding space.
A semi-supervised attribute network embedding approach is presented in \cite{huang2017label}.
Network embedding for social network with incomplete and noisy content information is proposed in \cite{zhang2017user}.
Many of these approaches are computationally expensive and are not scalable for very large datasets. Some of them (as observed in Section \ref{sec:exp}) also use content in a rigid way with structure such that the inconsistency present between the two sources affect the joint learning. So we propose a biased random walk based embedding approach which uses the informativeness of structure and content of a node, and learns the network representations by intelligently selecting the correct source at any phase of learning.


\section{Problem Statement}\label{sec:prob}
An information network is typically represented by a graph $G = (V, E, W, F)$, where $V=\{v_1, v_2,\cdots, v_n\}$ is the set of nodes (a.k.a. vertexes), each representing a data object. $E \subset \{(v_i,v_j) | v_i,v_j \in V \}$ is the set of edges between the vertexes. 
Each edge $e \in E$ is an ordered pair $e = (v_i, v_j)$ and is associated with a weight $w_{v_i,v_j} > 0$, which indicates the strength of the relation. $W$ is the set of all those weights. If $G$ is undirected, we have $(v_i, v_j) \equiv (v_j, v_i)$ and $w_{v_i,v_j} = w_{v_j,v_i}$; if $G$ is unweighted, $w_{v_i,v_j} = 1$, $\forall (v_i,v_j) \in E$.
$F = \{f_i \;|\; i \in \{1,2,\cdots,n\} \}$, where $f_i \in \mathbb{R}^d$ is the word vector (content) associated with the node $v_i \in V$. So $F$ can be considered as the content matrix. For simplicity we assume that the content associated with each node is only textual in nature. This can be generalized easily to other type of content such as image or videos.

%
Traditionally $F$ can be represented by bag-of-word models. In a bag-of-word model, 
typically stop words are removed, and stemming is done as a preprocessing step. Each row of this matrix is a tf-idf vector for the textual content at the corresponding node. So the dimension of the matrix $F$ is $n \times d$, where $d$ is the number of unique words (after the preprocessing) in the corpus.

Given $G$, the task is to find some low dimensional vectorial representation of $G$ which is consistent with both the structure of the network and the content of the nodes. More formally, for the given network $G$, the network embedding is to learn a function $f : v_i \mapsto \mathbf{y_i} \in \mathbb{R}^K$, i.e., it maps every vertex to a $K$ dimensional vector, where $K < min(n,d)$. The representations should preserve the underlying semantics of the network. Hence the nodes which are close to each other in terms of their positional distance or similarity in content should have similar representation. This representation should also be compact and continuous as that would help in designing the machine learning algorithms better .

\section{Solution Approaches}\label{sec:soln}
First we discuss an important preprocessing step that we use throughout the rest of the paper. Given the network $G$ with some content in the form of matrix $F$, our goal is to divide the network into two layers. The first layer corresponds to the structure, and the second layer is for the content of the network.\\

\textbf{Structure Layer}: Intuitively, this layer is the same as the given network without any content in the nodes. Mathematically, given the input network $G = (V, E, W, F)$ (as in Section \ref{sec:prob}), the structure layer is a graph $G_s = (V, E_s, W_s)$, with $E_s = E$ and $W_s=W$.\\

\textbf{Content Layer}: This layer is a directed graph which captures the similarity between pairs of nodes in terms of their respective contents. Again, given the input network $G$ as above, we define the content layer to be the graph $G_c = (V, E_c, W_c)$. Hence the nodes in the content layer are the same as the input graph. For each node $v_i \in V$, initially we compute the similarity or weight to all other nodes $v_j$ ($j \neq i$) based on the cosine similarity \cite{huang2008similarity} of the row vectors $F_{i.}$ and $F_{j.}$. So, $w_{ij}^c = Cosine(F_{i.}, F_{j.})$. But in this case, the content graph can be nearly complete (i.e., there is an edge between almost any pair of vertexes) and it would increase the computational time to process the graph. Whereas, in the structure layer, the number of edges is fixed. So we compute average number of outgoing edges over all the nodes in the Structure layer. Let's call it $avg_s$. If the given network is directed, $avg_s = |E|/n$, and if it is undirected then, $avg_s = 2 \times |E|/n$. We want the number of edges in the content layer to be comparable with the number of edges in the structure layer, so that it can help the random walk as discussed later. So in the content layer, for each node, we only retain the top
$\theta \times \lceil {avg_s}\rceil $ outgoing edges in terms of their edge weights, where $\theta$ is a positive integer.

Figure \ref{fig:flow} shows both structure and content layers, and their interconnections which we will discuss in Section \ref{sec:SaC2Vec}.


Next we propose some intuitive approaches to embed an information network with structure and content. These approaches are also based on biased random walk, as it has been shown to be effective and computationally efficient for network embedding. To make the paper self-contained, we brief the language model technique that our models would use.\\

\textbf{Language Modeling}: In any language modelling, the basic task is to maximize the likelihood of a sequence of words appearing in a document. If $W = (w_{0}, w_{1},..., w_{n})$ is a sequence, then the task is to maximize the probability of the next word conditioned on its context words. Mathematically, it can written in the following form:

\begin{equation}
\begin{aligned}
& \underset{w_n}{\text{maximize}}
& & {Pr}(w_{n}|w_{0}, w_{1},..,w_{n-1}) \\
\end{aligned}
\end{equation}

SkipGram can be used to generate embedding for words in vocabulary V \cite{mikolov2013distributed}. In SkipGram model, given a current word, it tries to predict the context. In DeepWalk paper, the SkipGram model is used beyond language modelling, to generate representation for nodes in a network. For a graph $G=(V,E)$ and for a vertex $v_{i}$, $\phi(v_{i})$ represents the embedding of the node. Considering $w$ being the window size of the node, the embeddings $\phi$ can be found by maximizing the following objective.
\begin{equation}
\begin{aligned}
& \underset{\phi}{\text{maximize}}
& & \log({Pr}\{ v_{i-w},\ldots, v_{i+w}\}\: \backslash \: v_{i} \: | \: \phi(v_{i}))\\
\end{aligned}
\end{equation}

\vskip3mm
Before going to the main algorithm SaC2Vec, we propose two simple and intuitive approaches to combine structure and content for network embedding.

\subsection{Convex Sum of Embeddings: CSoE}\label{sec:convex}

The idea behind the convex sum of embeddings is to generate embeddings independently for the graphs representing structure and content respectively using node2vec algorithm. As both the structure and content are important for a graph, we have taken a convex sum of the two embeddings. This method can be used only when the node's embedding dimension is same for structure and content.  \\
Let $e_s^i$ be the embedding of the node $v_i$ learned by using the node2vec algorithm on Structure graph $G_s = (V,E_s,W_s)$. Similarly we get the embedding $e_c^i$  for node $v_i$ using node2vec on Content graph $G_c = (V,E_c,W_c)$. The final embedding in this approach is generated using the convex combination of the embeddings $e_s^i$ and $e_c^i$ component-wise giving the same length embedding $e_{convex}^i$.\\
In this approach, if the dimension of the embedding of a node $v_i$ corresponding to the structure layer is $K_s$ and to the content layer is $K_c$, then $K_s = K_c = K$ and the embedding size for the same node after taking the convex sum will also be $K$.\\
Mathematically, the convex embedding for the $i^{th}$ node is:
\begin{equation}\label{eq:summed}
e_{convex}^{i} = \alpha . e_{s}^{i} + (1 - \alpha).e_{c}^{i} 
\end{equation}
\[e_{s}^{i} \in \mathbb{R}^{K} \:, \quad
e_{c}^{i} \in \mathbb{R}^{K} \:, \quad
e_{convex}^{i} \in \mathbb{R}^{K} \]
\[\quad \forall i \in  \{ 1,2,\ldots,n\} 
	\quad \textnormal{and} \quad 0 \leqslant \alpha \leqslant 1 \]



\subsection{Appended Embeddings: AE} \label{sec:appended}
In this approach, we append the embeddings $e_s^i$ and $e_c^i$ generated as mentioned in \ref{sec:convex} for node $v_i$ corresponding to both structure graph and content graph. In this approach the node's embedding dimension can be different for structure and content as opposed to previous method.\\
Hence, if the length of node embedding corresponding to a node $v_i$ of a structure is $K_s$ and for content is $K_c$, then the embedding size for the same node after appending the embeddings is ($K_s+K_c$).

\begin{equation}\label{eq:appended}
e_{appended}^{i} = [e_{s}^{i} || e_{c}^{i}] \\
\end{equation}
\[e_{s}^{i} \in \mathbb{R}^{K_s} \:, \quad
e_{c}^{i} \in \mathbb{R}^{K_c} \:, \quad
e_{appended}^{i} \in \mathbb{R}^{K_s+K_c} \]
\[\quad \forall i \in  \{ 1,2,\ldots,n\}  \]

\begin{figure*}[h]
\centering
\includegraphics[scale=0.28]{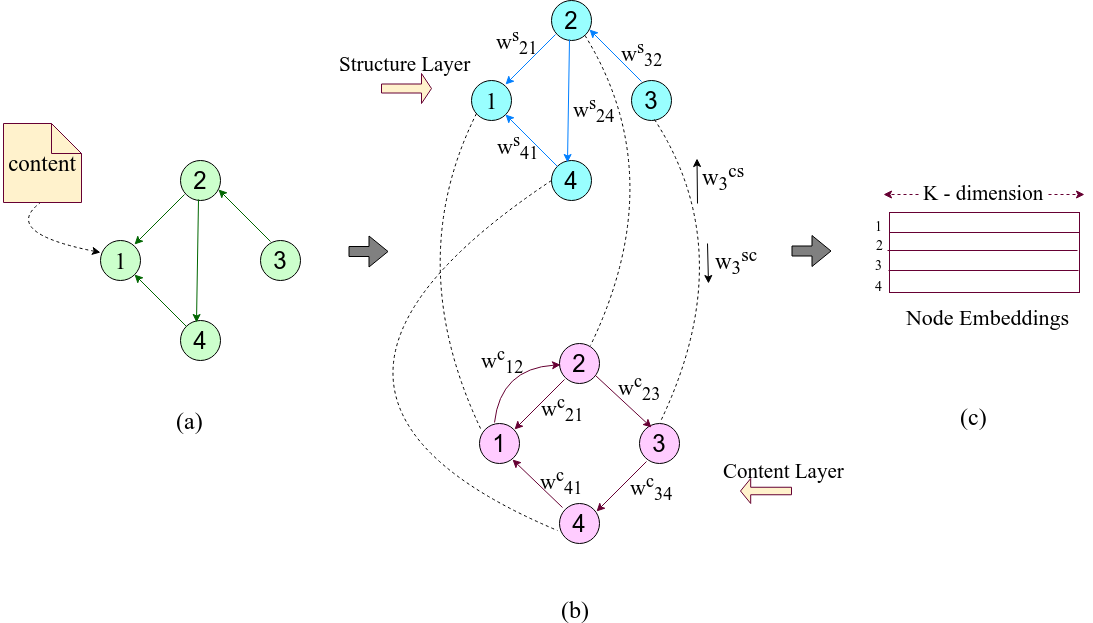}
\caption{This demonstrates the work flow of the proposed method SaC2Vec for the node embedding generation.
(a) Represents an information network with links between the nodes representing some relation. Also each node has some content associated with it. (b) Represent the two layered network. The upper network is the structure layer which can be (un)directed and (un)weighted depending on the input network. The lower network is the content network which is directed and weighted network formed using the method described in Section \ref{sec:soln}. There are 4 weighted and directed edges (shown by dashed line) connecting the corresponding nodes between the layers. (c) shows the K dimensional output embedding of the network.}
\label{fig:flow}
\end{figure*}

\vskip3mm
The above two approaches are simple, but they produce the embeddings from structure and content independently, and aggregate them as a post processing step to get the final embedding of the network. Ideally one should couple the embedding generation process from structure and content, and optimize them together. We discuss such an approach in the subsection below.

\begin{algorithm} 
  \caption{\textbf{SaC2Vec} - Structure and Content to Vector}
  \label{alg:sac2vec}
\begin{algorithmic}[1]
      
	\State Input: The network $G=(V,E,W,F)$, $K$: Dimension of the embedding space where $K<<min(n,d)$, $r$: Number of time to start random walk from each vertex, $l$: Length of each random walk 
    \State Output: The node embeddings of the network $G$
	\State Generate the structure layer and the content layer as described in Section \ref{sec:soln}
	\State Add the edges between the layers with weights as shown in Equations \ref{Eq:wsc} and \ref{Eq:wcs} to generate the multi-layered network
    \State $Corpus$ = [ ] \Comment{Initialize the corpus}
	\For{$iter \in \{1,2,\cdots,r\}$}
      \For{$v \in V$}
        \State select the node $v$ as the starting node of the random walk
        \State $Walk$ = [$v$] \Comment{Initialize the Walk(sequence of nodes)}
        \For{$walkIter \in \{1,2,\cdots,l\}$}
        	\State Select the layer to move next with probabilities as in \ref{Eq:ps} and \ref{Eq:pc}
            \State Move 1 step using node2vec to find the next node $v_i$ (Section \ref{sec:SaC2Vec})
            \State Append $v_i$ to $Walk$
        \EndFor
        \State Append Walk to Corpus
      \EndFor
    \EndFor
    \State Find the node embeddings by running language model on $Corpus$ (Section \ref{sec:soln})
	\end{algorithmic}
  \end{algorithm} 
  
\subsection{Solution Approach: SaC2Vec Model}\label{sec:SaC2Vec}

Here we propose \textit{SaC2Vec} model to embed an information network by coupling structure and content while learning the node representations. In this, we first get the structure layer and the content layer as discussed before. Then we connect the nodes which has one to one correspondence between the structure and the content layers. So, for any node $v_i$ in the original network, suppose the label of it in structure network is $v_i^s$ and the label in content layer is $v_i^c$. So we add two directed and weighted edges between $v_i^s$ and $v_i^c$ as follows.

Let us consider the directed edge $(v_i^s, v_i^c)$ with weight $w_i^{sc}$. To set the weight, we introduce few notations as described in \cite{ribeiro2017struc2vec}. Let us define the following for a node $v_i^s$ in the structure layer.
\begin{align}\label{eq:Gamma}
\Gamma_i^s = \{ (v_i^s, v_j^s) \in E_s \;|\; w_{(v_i^s, v_j^s)}^s \geq \frac{1}{|E_s|} \sum\limits_{e' \in E_s} w_{e'}^s \; , \; v_j^s \in V \}
\end{align}
It is basically the set of outgoing edges from $i$ whose weight is more than the average edge weight in the corresponding layer.
Similarly we can define $\Gamma_i^c$ for the nodes in the content layer. Now we set,
\begin{align}
w_i^{sc} = \ln(e + |\Gamma_i^s|)\label{Eq:wsc} 
\end{align}
Similarly we can associate a weight of $w_i^{cs}$ with the edge $(v_i^c, v_i^s)$, from content layer to the structure layer as follows. 
\begin{align}
w_i^{cs} = \ln(e + |\Gamma_i^c|)\label{Eq:wcs}
\end{align}
Hence, we have generated a multiplex or multi-layered graph using structure and content, as shown in Figure \ref{fig:flow}. Next we define a bias random walk on this multi-layered graph.

The intuition of the random walk is as follows. Given, at a particular time-step of the random walk, we are at node $v_i$, either in the structure or in the content layer. Before taking the next step, we first calculate the probability of taking that step either into the structure layer or into the content layer. Our goal is to move to a layer which is more informative in some sense at node $v_i$. Let us define the probabilities as:
\begin{align}
p(v_i^s | v_i) = \frac{w_i^{cs}}{w_i^{sc} + w_i^{cs}}\label{Eq:ps}\\
p(v_i^c | v_i) = 1 - p(v_i^s | v_i) = \frac{w_i^{sc}}{w_i^{sc} + w_i^{cs}}\label{Eq:pc}
\end{align}
Let us try to understand the probability of selecting the structure layer at node $v_i^s$. Clearly, larger the value of $w_i^{sc}$, higher the number of outgoing edges from the node $v_i^s$ with relatively high edge weights in the structure layer \cite{ribeiro2017struc2vec}. In that case, the random walk has many choices to move from the node $v_i^s$ if it remains to be in the structure layer (as discussed later). Where as, if the value of $w_i^{sc}$ is low, the random walk is likely to select only from a few nodes to move next. In the second case, the choice is more informative and less random. So when the value of $w_i^{sc}$ is high, we want to prefer content layer, and similarly when the value of $w_i^{cs}$ is high, we want to prefer structure layer.

\begin{figure}
\includegraphics[scale=0.20]{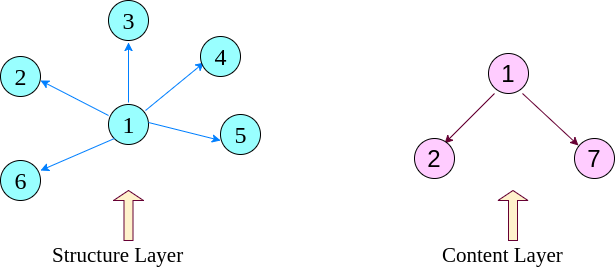}
\caption{Example to clarify the switching layer probabilities in SaC2Vec model}
\label{fig:intuition}
\end{figure}

We give an example to clarify this further. Figure \ref{fig:intuition} depicts a node $v_1$ having different number of edges (we have only shown the edges whose weights are more than the average and hence part of $\Gamma_1^s$ or $\Gamma_1^c$) from the vertex in the structure and the content layers respectively.
Say at any time step of the random walk, we are at node $v_1$. Now we need to choose the next node in the random walk. If we choose the next node on the basis of neighbors of $v_1$ in the structure layer then we will have 5 options to choose from, whereas if we choose neighbors of $v_1$ in content layer we only have 2 options to choose from.
Thus we can make more informed choice if we go to the content layer and then choose from the neighbors of $v_1$ in the content layer (i.e. perform 1 step of node2vec or random-walk from node $v_1$ in content graph). This justifies the probabilities in Eq. \ref{Eq:ps} and \ref{Eq:pc} of selecting the layer to move next in the random walk. Once a particular layer is selected, there is no role of the other layer in selecting the next vertex to visit in random walk.

Now we discuss the probability of selecting the next vertex from the current vertex $v_i$, given that we have selected a particular layer (either structure or content, depending on which $v_i$ would be either $v_i^s$ or $v_i^c$), as discussed above. We run \textit{one step} of node2vec algorithm \cite{grover2016node2vec} from the node $v_i$ in the selected layer. It is important to mention that, the node (suppose it was $v_j$, $j \neq i$) that we visited in the last step of node2vec, may not have a direct edge to the present node, in case if we have changed the layer. For example, we visited $v_i^s$ from $v_j^s$, and then changed from structure to content layer, where there is no direct edge from $v_j^c$ to $v_i^c$. In that case, the $p$ and $q$ parameter of node2vec will have no role to play and we will sample the next node only on the basis of the weighted sampling on the nodes directly connected to $v_i$. Otherwise, $v_j$ would act as the last node visited in the node2vec algorithm\footnote{Please check Section 3.2 in \cite{grover2016node2vec}}. 
\begin{table*}

    \caption{Summary of the datasets used: All the datasets used have both network structure and textual content in each node. \textit{\#Unique Words} counts the number of distinct words in a network. \textit{Class Distribution} is the proportion of different communities in a dataset. \textit{Inter/intra links} is the ratio between the number of inter community links to that of the intra community links in a dataset.}
	\centering
	\begin{tabular}{*7c}
	\toprule
	\sffamily{Dataset} & \#Nodes & \#Edges & \#Labels & \#Unique & Class & Inter/Intra \\
    \sffamily{} & & & & Words & Distribution & links \\
    \hline
	\midrule
    \sffamily{Cora} & 2708 & 5429 & 7 & 1433 & 11:15:30:16:8:7:13 & 0.22 \\
    \sffamily{Citeseer} & 3312 & 4715 & 6 & 3703 & 18:8:21:20:18:15 & 0.34 \\
    \sffamily{Flickr} & 7575 & 239738  & 9 & 12047 & 11:10:11:11:10:12:11:12:12 & 3.19 \\
    \sffamily{Pubmed} & 19717 & 44338 & 3 & 500& 21:40:39 & 0.25\\
\bottomrule
	\end{tabular}
	\label{tab:data}
	\end{table*} 

The next node selected by the node2vec step is added to the sequence of the random walk. We repeat the above step $l$ times, where $l$ is the length of the random walk that we want to generate from each node. Algorithm \ref{alg:sac2vec} summarizes the whole process.
\begin{lemma}
Asymptotic time complexity of SaC2Vec is $O((|V| + |E|) \log{|V|})$, which is same as that of node2vec.
\end{lemma}
\begin{proof}
Here we aim to prove, even though we are able to use content with structure, the asymptotic time complexity of SaC2Vec is still the same as node2vec. If we assume that the number of random walks from each node and the length of each random walk are constant, then the time complexity of node2vec is $O((|V| + |E|) \log{|V|})$, due to the use of alias table to compute the transition probabilities.
The time complexity of the random walk in \textit{SaC2Vec} is similar to that of the node2vec algorithm. The only overhead here is the construction of the content network which in worst case can $O(|V|^2)$. But since we want the number of edges for each node in the content network to be in the order of the average number of edges in reference graph, the content graph will have very less number of edges as compared to a complete graph. This fact can be exploited to reduce the overhead time for content network creation using an approach similar to Space Partitioning Tree \cite{shen1999fast}.

Space Partitioning Tree(SPT) is a data structure that allows us to find the closest object to another object in logarithmic time. But it can only operate under certain distance functions which satisfy non-negativity, identity of indiscernibles, symmetry and triangle inequality properties.
As cosine similarity does not satisfy the non-negativity property, so we can not use the SPT straight away. If we add 1 to the cosine similarity value (which ranges from -1 to 1), then the first three properties are satisfied but the triangle equality property is not. It can be shown that, if we use normalized feature vectors then the cosine ranking is equivalent to euclidean ranking and the relationship between them can be described by $||x-y|| = \sqrt[]{2 - 2\cos(x,y)}$, where $x$ and $y$ are two normalized vectors.
Euclidean distance being a valid metric for SPT, we can also use the normalized cosine ranking for SPT and thus k-nearest neighbors for a node can be calculated in $O(\log{|V|})$ time. Hence, the entire content graph can be formed in $O(|V|log{|V|})$ time.
So the total time complexity of Sac2Vec is $O(|V|log{|V|} + (|V| + |E|) \log{|V|}) = O((|V| + |E|) \log{|V|})$.
\end{proof}



\section{Generalization to Multiple Content Types}\label{sec:generalization}
Different social networks now-a-days contain multiple types of content such as rich textual information, images, videos, etc., along with the link structure. Common examples can be a set of user profiles in Facebook where users post text data, images and videos, a set of linked video pages in Youtube where each page contains description of the video along with the video itself. Trivially one can concatenate or aggregate the features from all the different sources of content, and treat them equally. But this is definitely not the best way, as these sources of content are very different in nature. So in this section, we extend the proposed SaC2Vec from a single source of content to multiple sources of content.

\begin{figure*}[h]
\centering
\includegraphics[scale=0.28]{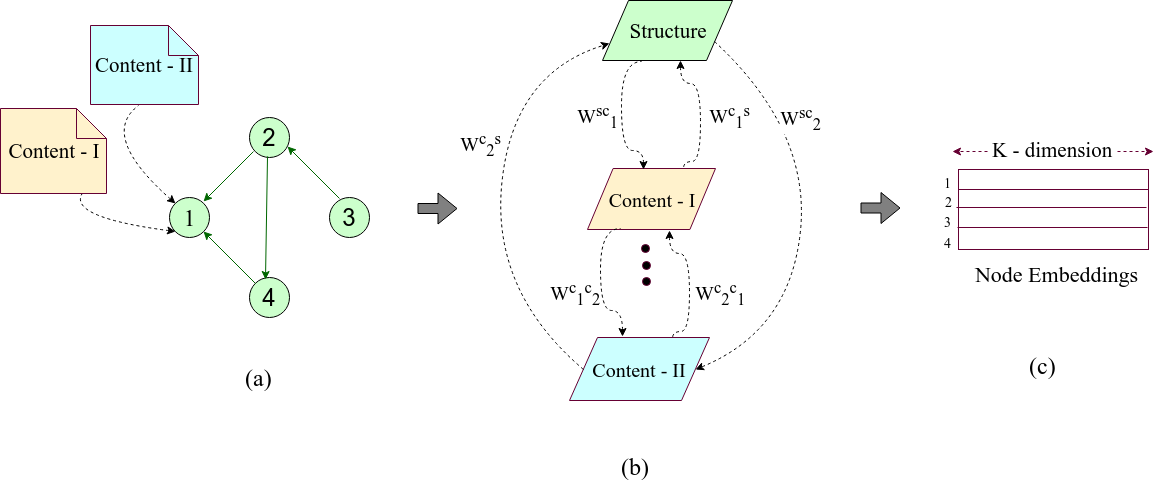}
\caption{Generalization of SaC2Vec to the case when there are multiple types of contents associated with each node. We first make a multiplex network having one layer for the link structure of the network, and one layer for each type of content.}
\label{fig:multilayer}
\end{figure*}

The extension is simple. We again create a multiplex network from the given information network with different types of content associated with each node. Similar to Section \ref{sec:SaC2Vec}, there is one layer corresponding to the link structure of the network - let us call that structure layer as before. Then we can compute one layer for each type of content. Let us use the index variable $t$ to refer to these layers. Without loss of generality, assume $t \in \{1,2,\cdots,T\}$ where $1$ stands for the structure layer and rest are for the content layers. For each node in the given network, we create one node in each layer of this multiplex network, and connect the corresponding nodes between any two layers. Link structure within the structure layer is same as the input network, whereas the link structure within each content layer can be formed in the same way as described in Section \ref{sec:soln}. Figure \ref{fig:multilayer} gives a high level overview of the idea. To connect the nodes between different layers, we can again define $\Gamma_i^t$ as in equation \ref{eq:Gamma}, $\forall t = 1,2,\cdots, T$. Let us also define the weight of the edge $(v_i^t, v_i^{t'})$ as $w_i^{tt'} = \log(e + |\Gamma_i^t|)$, $t \neq t'$. Clearly, all the outgoing edges from $v_i^t$ has the same weight.

Again similar to the case of single content type, at a particular time step of the random walk when we are at node $v_i$ in one of the $T$ layers, before the next move, there is always a chance for the random walker to go to a different layer or to remain at the same layer. For the two layer case, from Eq. \ref{Eq:pc} we know that $p(v_i^s | v_i) = \frac{w_i^{cs}}{w_i^{cs} + w_i^{sc}}$. Extending that to multiple layer case, the probability that the random walk would go to the layer $t$, given it is there at node $v_i$ at any layer, can be computed as:
\begin{equation}\label{eq:probMultiContent}
p(v_i^t | v_i) = \frac{\mathlarger{\sum}\limits_{t^{'} : t^{'} \neq t} w_i^{t't}}{\mathlarger{\sum}\limits_{t^{'}, t^{''} : t^{'} \neq t^{''}} w_i^{t^{'}t^{''}}}
\end{equation}
The intuition of selecting the layer is similar to the two layer case. We want to prefer a layer where there are less number of heavy weight edges, so that the next move of the random walk is less random and more informative. Once we selected the layer to find the next vertex in the random walk, we can use one step of node2vec in the same way we have used in Section \ref{sec:SaC2Vec}. This process of selection of layer and moving to the next vertex will be followed until the length of the random walk is reached. After that, we can use the same language modeling technique to find the embedding of the nodes in the given network.

\section{Experimental Evaluation}\label{sec:exp}
In this section, we experimentally evaluate the performance of the proposed algorithms and compare the results with state-of-the-art network embedding algorithms. We used 4 real life publicly available network datasets which contain both link structure and the content of each node. To evaluate the quality of the generated embeddings, we have selected three different types of machine learning based network mining tasks: node classification, node clustering and network visualization. We find that our proposed algorithms (specially SaC2Vec) outperform all the state-of-the-art approaches across different learning tasks. For simplicity, we have considered the datasets having only a single type of content. As the datasets have only one type of content here, evaluation on multiple types of content (as stated in \ref{sec:generalization}) will be done in the future.

\subsection{Datasets Used}
The following datasets have been used in the paper:

\textbf{Cora}\footnote{\label{footnote:dataurl}\url{https://linqs.soe.ucsc.edu/data}}: The Cora dataset consists of Machine Learning papers. The citation between them forms the network. Each paper attribute consists of 1433 unique words.

\textbf{Citeseer}\textsuperscript{\ref{footnote:dataurl}}:  It is a citation network consisting of 3312 scientific publication. Each publication attribute consists of 3703 unique words.

\textbf{Flickr}\footnote{\url{https://github.com/xhuang31/AANE_Python/blob/master/Flickr.mat}} is an online community where people can share photograph and can also follow each other which form a network. The tags specified on the image act as the attribute. The group which the photographer has joined acts its label.

\textbf{Pubmed}\textsuperscript{\ref{footnote:dataurl}}: It consists of scientific publications from PubMed database related to diabetes classified into one of three classes: Experimental, Type1 or Type2. The content consists of 500 unique words.

A detailed description of dataset is provided in Table \ref{tab:data}
  
\subsection{Baseline Algorithms and Experimental Setup}
We compare the performance of our algorithms against the following state-of-the-art approaches: DeepWalk \cite{perozzi2014deepwalk}, LINE \cite{tang2015line}, node2vec \cite{grover2016node2vec}, TADW \cite{yang2015network}, AANE \cite{huang2017accelerated} and GraphSAGE \cite{hamilton2017inductive}. Among these baselines, DeepWalk, LINE and node2vec use only network structure, whereas TADW, AANE and GraphSAGE use both structure and the content of the nodes for generating the embeddings.
We create two layers, one each for structure and content, from the input graph for our proposed algorithms, we also run DeepWalk, LINE and node2vec independently on these two layers to see the difference of using only structure and that with only content for these three baselines.
We have mostly used the default settings of parameter values present in the publicly available implementations of the respective baseline algorithms. We used GraphSAGE in a non-inductive setting, i.e., all the nodes in the network were present from the beginning, for the sake of fair comparison.

In our experiments, two proposed intuitive approaches CSoE and AE, as well as the proposed SaC2Vec are evaluated for multi-class node classification, node clustering and network visualization. We have kept the embedding dimension for all the datasets to be 128 for SaC2Vec.

\begin{table}[!htbp]
\caption{Performance of the algorithms for Classification with Random Forest on the Citeseer data for different embedding methods. We have tried the methods using both structure and content graph. Result for DeepWalk Struct. represent the accuracy of the classifier trained on embeddings learned using the structure network for the corresponding dataset. Similarly Deepwalk Cont. represents the accuracy of the classifier trained on embeddings learned using the content layer. TADW, AANE, GraphSAGE, CSoE, AE and SaC2Vec use both structure and content for learning the network embedding.}
\begin{center}
\resizebox{\columnwidth}{!}{%
\begin{tabular}{c|cc|ccccc}
    \toprule
    \multirow{2}{*}{Metric} & \multicolumn{2}{c}{\multirow{2}{*}{Algorithm}} &  \multicolumn{5}{|c}{Train Size(\%)} \\
    & & & 10 & 20 & 30 & 40 & 50 \\
    \hline
    \hline
    \multirow{10}{*}{Macro-F1} & \multirow{2}{*}{DeepWalk} & Struct. & 48.23 & 55.11 & 59.66 	& 63.07 & 64.97\\[1mm]
    & & Cont. & 42.52 & 50.38 & 55.69 & 57.99 & 60.10   \\[2mm]
   
   & \multirow{2}{*}{LINE} & Struct.  & 19.84 & 21.61 & 22.72 & 24.86 & 25.14        \\[1mm]
    & & Cont. & 30.34 & 35.43 & 38.23 & 40.80 & 42.09            \\[2mm]
   
   & \multirow{2}{*}{node2vec} & Struct.  & 50.86 & 56.52 & 59.69 & 62.51 & 63.84     \\[1mm]
    & & Cont. & 40.12 & 48.48 & 53.77 & 56.32 & 58.78    \\[2mm]
    & TADW & &  56.80 & 62.59 & 65.86  & \textbf{67.21} & \textbf{68.20}       \\[2mm]
    & AANE & &  49.51 & 51.47 & 53.73 & 54.22 & 55.01       \\[2mm]
    & GraphSAGE & & 30.91 & 37.58 & 40.33 & 43.57 & 44.60       \\[2mm]
    & \textbf{CSoE} &  & 48.60 & 55.08 & 58.43 & 61.56 & 64.62       \\[2mm]    
	& \textbf{AE} & & 51.29 & 57.49 & 61.32 & 64.61 & 65.59      \\[2mm]
    & \textbf{SaC2Vec} & &  \textbf{62.14} & \textbf{64.19} & \textbf{65.89} & 67.20 & 68.18      \\ [2mm]
     
     \hline 
 
    \multirow{10}{*}{Micro-F1} & \multirow{2}{*}{DeepWalk} & Struct. & 53.29 & 59.24 & 63.73 		& 67.04 & 68.78\\[1mm]
    & & Cont.  & 47.28 & 55.07 & 60.13 & 62.53 & 64.43   \\[2mm]
    
    & \multirow{2}{*}{LINE} & Struct.  & 24.42 & 26.07 & 26.87 & 28.84 & 28.94   \\[1mm]
    & & Cont. & 37.38 & 42.18 & 45.35 & 47.89 & 49.25             \\[2mm]
    
    & \multirow{2}{*}{node2vec} & Struct.  & 55.84 & 60.98 & 63.88 & 66.41 & 67.71     \\[1mm]
    & & Cont. & 44.13 & 52.11 & 57.37 & 59.64 & 62.46  \\[2mm]
    & TADW & &  63.61 & 68.46 & 71.44 & 72.50 & 73.24         \\[2mm]
    & AANE & &  56.48 & 59.50 & 61.24 & 61.90 & 62.39        \\[2mm]
    & GraphSAGE & & 38.01 & 44.34 & 46.82 & 49.47 & 50.34        \\[2mm]
    & \textbf{CSoE} &  &  54.33 & 59.93 & 63.30 & 65.55 & 68.72      \\[2mm]    
	& \textbf{AE} & &   57.48 & 62.29 & 65.56 & 68.93 & 69.40    \\[2mm]
    & \textbf{SaC2Vec} & &  \textbf{ 69.10} & \textbf{71.84 }& \textbf{73.07} &\textbf{ 74.01} & \textbf{74.61}     \\ [2mm]
    
    \bottomrule
 \end{tabular}
}
\end{center}
\label{tab:classification}
\end{table}

\begin{table}[!htbp]
\caption{Performance of the algorithms for Classification on the Flickr data with same setting as in Table \ref{tab:classification}.}
\begin{center}
\resizebox{\columnwidth}{!}{%
\begin{tabular}{c|cc|ccccc}
    \toprule
    \multirow{2}{*}{Metric} & \multicolumn{2}{c}{\multirow{2}{*}{Algorithm}} &  \multicolumn{5}{|c}{Train Size(\%)} \\
    & & & 10 & 20 & 30 & 40 & 50 \\
    \hline
    \hline
    \multirow{10}{*}{Macro-F1} & \multirow{2}{*}{DeepWalk} & Struct. & 37.98 & 40.54 & 42.03 & 42.43 & 43.50\\[1mm]
    & & Cont. & 73.71 & 77.45 & 78.71 & 79.49 & 80.01   \\[2mm]
   
   & \multirow{2}{*}{LINE} & Struct.  & 25.52 & 27.72 & 28.67 & 29.38 & 30.03       \\[1mm]
    & & Cont. & 51.45 & 58.78 & 61.98 & 63.68 & 65.38           \\[2mm]
   
   & \multirow{2}{*}{node2vec} & Struct.  & 43.69 & 46.02 & 47.26 & 48.38 & 49.37     \\[1mm]
    & & Cont. & 77.69 & 80.28 & 81.68 & 82.55 & 82.75   \\[2mm]
    
    & TADW & &  74.99 & 79.06 & 80.26 & 81.55 & 82.35     \\[2mm]
    & AANE & &  71.31 & 72.93 & 75.04 & 75.67 & 76.37      \\[2mm]
    & GraphSAGE & & 17.52 & 19.37 & 19.79 & 20.82 & 21.21       \\[2mm]
    & \textbf{CSoE} &  & 77.99 & 80.01 & 81.62 & 82.45 & 82.76      \\[2mm]    
	& \textbf{AE} & & \textbf{78.77} & \textbf{81.50} & \textbf{83.12} & \textbf{83.86} & \textbf{84.36 }   \\[2mm]
    & \textbf{SaC2Vec} & & 78.36 & 80.84 & 82.17 & 82.86 & 82.93    \\ [2mm]
     
     \hline 
 
    \multirow{10}{*}{Micro-F1} & \multirow{2}{*}{DeepWalk} & Struct. & 39.52 & 42.04 & 43.49 & 44.07 & 44.96\\[1mm]
    & & Cont.  &74.29 & 77.88 & 79.09 & 79.87 & 80.29   \\[2mm]
    
    & \multirow{2}{*}{LINE} & Struct.  & 26.74 & 28.87 & 29.78 & 30.52 & 31.26  \\[1mm]
    & & Cont. & 52.43 & 59.23 & 62.44 & 64.22 & 65.85             \\[2mm]
    
    & \multirow{2}{*}{node2vec} & Struct.  & 45.37 & 47.26 & 48.41 & 49.42 & 50.47     \\[1mm]
    & & Cont. & 78.15 & 80.73 & 82.17 & 83.12 & 83.19\\[2mm]
    & TADW & &  75.25 & 79.12 & 80.38 & 81.65 & 82.51        \\[2mm]
    & AANE & &  71.76 & 73.23 & 75.26 & 75.92 & 76.55       \\[2mm]
    & GraphSAGE & & 19.00 & 20.47 & 20.63 & 21.88 & 22.39        \\[2mm]
    & \textbf{CSoE} &  &  78.51 & 80.62 & 82.15 & 82.96 & 83.21     \\[2mm]    
	& \textbf{AE} & &  \textbf{79.36} & \textbf{82.00} & \textbf{83.51} &\textbf{ 84.04} & \textbf{84.70}   \\[2mm]
    & \textbf{SaC2Vec} & &  78.81 & 81.20 & 82.41 & 82.72 & 83.19    \\ [2mm]
    
    \bottomrule
 \end{tabular}
}
\end{center}
\label{tab:classification2}
\end{table}

\subsection{Multi-class Node Classification}
Node classification is an important application useful in cases when labeling information is available only for a small subset of nodes in the network. This information can be used to enhance the accuracy of the label prediction task on the remaining/unlabeled nodes. 
For this task, firstly we get the embedding representations of the nodes and take them as the features to train a random forest classifier \cite{liaw2002classification}.
We split the set of nodes of the graph into training set and testing set. The training set size is varied from 10\% to 50\% of the entire data. The remaining (test) data is used to compare the performance of different algorithms.
We take two popularly used evaluation criteria based on F1-score, i.e., Macro-F1 and Micro-F1 to measure the performance of multi-class classification algorithms. Micro-F1 is a weighted average of F1-score over
all different class labels. Macro-F1 is an arithmetic average of F1-scores of all output class labels. Normally, the higher the values are, the better the classification performance is. We repeat each experiment 10 times and report the average results.


The results for multi-class classification task on Citeseer dataset are presented in Table \ref{tab:classification}. First, we observe that the performance of DeepWalk and node2vec on the structure layer is better than that on the content layer, whereas for LINE it is just the reverse. Hence it is not possible to conclude which layer is more informative in general for this dataset. Whereas from the results for Flickr dataset, presented in Table \ref{tab:classification2}, it is clearly visible that the performance of DeepWalk, LINE and node2vec is significantly better on the content layer than the structure layer. Hence we can conclude that content layer is more informative for Flickr dataset.

It turns out that for Citeseer dataset, SaC2Vec performs the best in terms of Micro-F1 for all the training sizes. With Macro-F1 score, SaC2Vec performs the best when the training size is less than 40\%, and TADW performs slightly better (with a margin of 0.01\%-0.02\%) when training size is 40\%-50\%. In case of Flickr dataset, AE performs best for all the training sizes, both in Micro-F1 and Macro-F1 scores, beating SaCVec marginally (0.5\%-1.5\%).  Hence the proposed algorithm SaC2Vec is able to learn well even when the available labeled data is very small. Surprisingly, even the two intuitive and simple methods, CSoE and AE, proposed by us are also able to perform good 
compared to many state-of-the-art approaches on this dataset. We can also see that AE is performing better than CSoE as the embedding features are preserved in AE while they are lost in CSoE due to the convex combination.

\begin{table*}[ht]
\caption{Clustering Accuracy for different dataset and different embedding methods using KMeans++ algorithm. Accuracy corresponding to Struct. for a dataset and embedding method implies only structure layer is used for the embedding generation. Similarly Cont. represents the corresponding dataset and embedding method using only content layer. Interestingly, SaC2Vec is able to identify and exploit more from the better layer automatically during the learning process.}
\begin{center}
\begin{tabular}{ccccccccccccc}
    \toprule
    \multirow{1}{*}{Dataset} & \multicolumn{2}{c}{DeepWalk} & \multicolumn{2}{c}{LINE} & \multicolumn{2}{c}{node2vec} & \multicolumn{1}{c}{TADW} & \multicolumn{1}{c}{AANE} & \multicolumn{1}{c}{GraphSAGE} & \multicolumn{1}{c}{\textbf{CSoE}} & \multicolumn{1}{c}{\textbf{AE}} & \multicolumn{1}{c}{\textbf{SaC2Vec}} \\
    
    & \multicolumn{1}{c}{Struct.} & \multicolumn{1}{c}{Cont.} & \multicolumn{1}{c}{Struct.} & \multicolumn{1}{c}{Cont.} & \multicolumn{1}{c}{Struct.} & \multicolumn{1}{c}{Cont.} \\
    
    \midrule
    Cora &59.9 &56.9 & 20.86 & 31.57 &62.6 &29.2 &61.6 &  33.45 &  28.32 & 59.6 & 59.6 &\textbf{68.2}\\
    Citeseer & 44.4 & 31.4 & 21.86 & 39.28 & 44.1 & 54.4 & 65.5 & 39.82 & 30.89 & 52.5 & 51.6 &\textbf{69.5}\\
    Flickr & 31.20 & 73.92 & 25.33 & 81.09 & 31.06 & 75.45 & 25.12 & 28.95 & 18.05 & 75.76 & 77.18 & \textbf{85.9} \\
    Pubmed & 67.1 & 60.4 & 35.68 & 40.43 & 68.3 & 38.5 & 48.32 &   46.25 & 55.79 & 68.9 & \textbf{69.2} & 68.3\\

    \bottomrule
\end{tabular}
\end{center}
\label{tab:cluster}
\end{table*}

\subsection{Node Clustering}
Node Clustering is an unsupervised method of grouping the nodes into multiple communities or clusters.
First we run all the embedding algorithms to generate the embeddings of the nodes.
We use the node's embedding as the features for the node and then apply KMeans++ algorithm which is a modification of the KMeans with only difference in the initialization of the cluster centers \cite{kmeans++}. Since we are performing the clustering in a totally unsupervised setting, KMeans++ just divides the data into different classes. To find the test accuracy we need to assign the clusters with an appropriate label and compare with the ground truth labeling which is available for each dataset that we have considered. For finding the test accuracy we use unsupervised clustering accuracy \cite{xie2016unsupervised} which uses different permutations of the labels and choose the label ordering which gives best possible accuracy. Mathematically,
\begin{align}
Acc(\mathcal{\hat{C}},\mathcal{C}) = \max_{\mathcal{P}} \frac{ \sum\limits_{i=1}^n \mathbf{1}(\mathcal{P}(\mathcal{\hat{C}}_i)  = \mathcal{C}_i)) }{n}
\end{align}
Here $\mathcal{C}$ is the ground truth labeling of the dataset such that $\mathcal{C}_i$ gives the ground truth label of $i$th data point. Similarly $\mathcal{\hat{C}}$ is the clustering assignments discovered by some algorithm, and $\mathcal{P}$ is a permutation on the set of labels.
We assume $\mathbf{1}$ to be a logical operator which returns 1 when the argument is true, and otherwise returns 0.

The clustering accuracies for different datasets and algorithm are shown in Table \ref{tab:cluster}. One can see that the algorithms proposed by us outperform all the state-of-the-art algorithms. It turns out that for the datasets Cora and Citeseer, algorithms which use only one of structure or content are not able to perform well. TADW was the best among the baseline algorithms. But SaC2Vec is the most successful algorithm in combining structure and content to produce the embeddings. In case of Flickr dataset, TADW, AANE and GraphSAGE performs badly. Although node2vec, LINE and Deepwalk are showing good results on the content layer, SaC2Vec is performing best. 
For the Pubmed dataset, TADW suffers badly, but DeepWalk and node2vec on the structure layer for this dataset perform very well. Interestingly, the performance of SaC2Vec is same as the performance of node2vec applied only on the structure layer (without the content). It means SaC2Vec is able to understand the possible inconsistency of the content layer during the learning process and embeddings were learnt mostly from the structure layer. Thus SaC2Vec is a robust algorithm and less prone to noise. Overall, Appended Embedding (proposed in Section \ref{sec:appended}) performs the best on Pubmed, as the final embedding is just the concatenation of the embeddings found from the two layers, and thus embeddings from structure are still able to separate the respective communities. Overall, while other algorithms could not perform consistently over the different datasets, SaC2Vec is able to perform well for all the datasets and always turns out to be the best or very close to the best among the other approaches.

    

\subsection{Network Visualization}
In network visualization, the whole network is projected into a 2D space, and the goal is to  project it in such a way that the nodes in the same communities are placed close to each other, and nodes from different communities are placed far apart in the 2D space.
Network visualization is unsupervised as the labels of the nodes are not being used for learning the map, they can be used in the 2D plots for a better understanding of the quality of visualization. 
Again we first run the embedding algorithms to generate embedding for each node in a network.
We use ISOMAP \cite{tenenbaum2000global} toolkit present in python Scikit-learn \cite{pedregosa2011scikit} library to convert these embeddings to 2D space. We use same color for the nodes which belong to the same community, and different colors for the different communities. 

\begin{figure*}[h]
\centering
\begin{subfigure}{0.18\textwidth}
\includegraphics[scale=0.16]{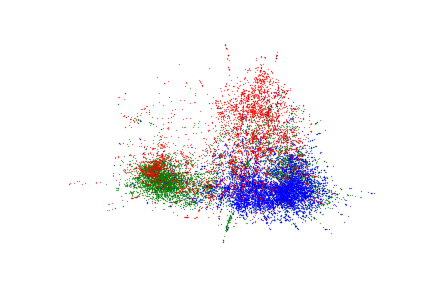}
\caption{DeepWalk}
\label{fig:subim1}
\end{subfigure}
\begin{subfigure}{0.18\textwidth}
\includegraphics[scale=0.16]{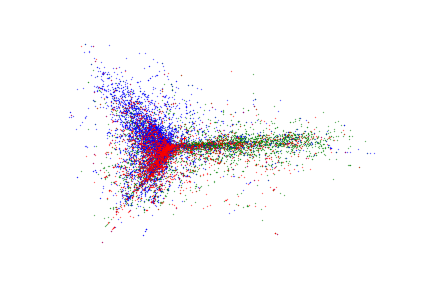}
\caption{LINE}
\label{fig:subim2}
\end{subfigure}
\begin{subfigure}{0.18\textwidth}
\includegraphics[scale=0.16]{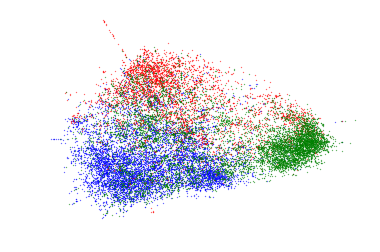}
\caption{node2vec}
\label{fig:subim3}
\end{subfigure}
\begin{subfigure}{0.18\textwidth}
\includegraphics[scale=0.16]{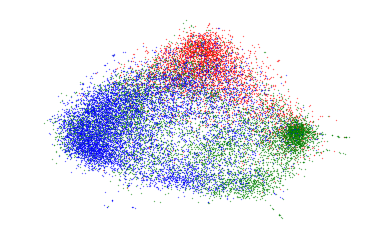}
\caption{GraphSAGE}
\label{fig:subim4}
\end{subfigure}
\\
\begin{subfigure}{0.18\textwidth}
\includegraphics[scale=0.16]{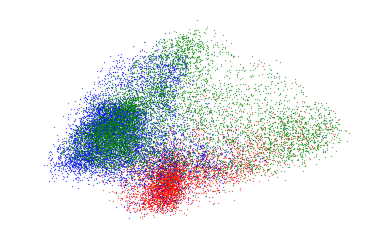}
\caption{AANE}
\label{fig:subim4}
\end{subfigure}
\begin{subfigure}{0.18\textwidth}
\includegraphics[scale=0.16]{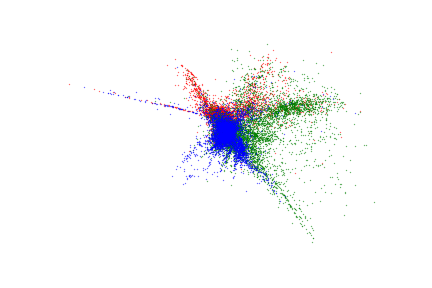}
\caption{TADW}
\label{fig:subim4}
\end{subfigure}
\begin{subfigure}{0.18\textwidth}
\includegraphics[scale=0.16]{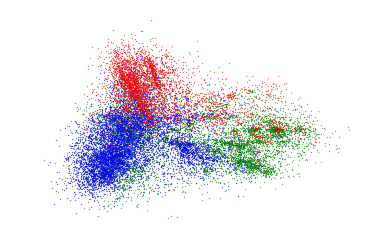}
\caption{\textbf{SaC2Vec}}
\label{fig:subim4}
\end{subfigure}
\caption{Visualization for the Pubmed dataset for different embedding methods. We have used ISOMAP algorithm to reduce the embedding to 2-D space. The nodes belonging to different classes are plotted using different colors.}
\label{fig:visual}
\end{figure*}

We have presented the results of visualization in Figure \ref{fig:visual}. The Pubmed dataset has three classes. Hence, the visualization plots have three clusters for all the algorithms and they are represented by three different colors. For DeepWalk, node2vec and LINE, we have considered their performance only on the structure layer as the plots for the the content layer were not good in general. It can be seen that SaC2Vec is able to discriminate the nodes based on their communities. Surprisingly, TADW and AANE, though they use both structure and content, are not able to discriminate the classes in the 2D space. We also observe that node2vec and GraphSAGE are able to produce descent visualization of the network and close to that of SaC2Vec.


\section{Discussion and Future Work}\label{sec:con}
In this work, we motivate the use of content along with the topological structure of an information network to generate the node embeddings. We first proposed some simple and intuitive algorithms which use both structure and content, and then proposed a novel algorithm \textit{SaC2Vec} to embed an information network by creating multi-layered network and then employ random walk for learning the network representation. We extend the SaC2Vec model for a scenario where each node can have different types of content such as text, images, etc. in it. Through experimentation we show that the embedding found can be used successfully for different network mining tasks. Comparison of the results with the six state-of-the-art algorithms shows the usefulness of our approach for network embedding. Experiments also show the robustness of SaC2Vec in the sense that it is able to intelligently select structure or the content in the case when one of them is noisy or inconsistent.

In the present experimentation, we have used datasets which have only one type of content along with the structure information. In the future, we want to conduct experiments with datasets having different types of content.
Also in this work, we have considered only the static graphs i.e. the network structure is time-independent. In contrast to static networks, dynamic networks are those that change over time and usually encountered in real-life situations. In the future work, we propose to extend our idea to the area of dynamic networks.

\bibliographystyle{ACM-Reference-Format}
\bibliography{SaC2Vec}

\end{document}